\documentclass[preprint,5p]{elsarticle}

\usepackage{graphicx}
\usepackage{dcolumn}
\usepackage{bm}
\usepackage{underscore}
\usepackage{color,xcolor}
\usepackage{amsmath}
\usepackage{amssymb}
\usepackage{amsfonts}
\usepackage{autobreak}
\usepackage{breqn}
\usepackage{caption}
\pdfoutput=1

\journal{Physics Letters A}

\begin{document}

\begin{frontmatter}



\title{Squeezed driving induced entanglement and squeezing among cavity modes and magnon mode in a magnon-cavity QED system}


\author[mymainaddress,mysecondaryaddress]{Ying Zhou}
\author[mymainaddress]{Jingping Xu\corref{mycorrespondingauthor}}
\cortext[mycorrespondingauthor]{Corresponding author}
\ead{xx\_ jj\_ pp@tongji.edu.cn}
\author[mymainaddress]{Shuangyuan Xie\corref{mycorrespondingauthor}}
\ead{xieshuangyuan@tongji.edu.cn}
\author[mymainaddress]{Yaping Yang}

\address[mymainaddress]{MOE Key Laboratory of Advanced Micro-Structured Materials, School of Physics Science and Engineering, Tongji University, Shanghai 200092, China}

\address[mysecondaryaddress]{School of Electronics and Information Engineering, Taizhou University, Taizhou, Zhejiang, 318000, China}

\begin{abstract}
We propose a scheme to generate entanglement between two cavity modes and squeeze magnon mode in a magnon-cavity QED system, where the two microwave cavity modes are coupled with a massive yttrium iron garnet (YIG) sphere through magnetic dipole interaction. The nonlinearity used in our system originates from a squeezed driving via parametric down-conversion process, which is the reason to cause entanglement and squeezing. By using the mean field approximation and employing experimentally feasible parameters, we demonstrate that the system shows zero entanglement and squeezing without squeezed driving. Meanwhile, our QED system denotes that the entanglement between squeezed cavity mode and magnon mode can be transferred to the other cavity mode and magnon mode via magnon-cavity coupling interaction, and then the two cavity modes get entangled. A genuinely tripartite entangled state is formed. We also show that magnon mode can be prepared in a squeezed state via magnon-cavity beam-splitter interaction, which is as a result of the squeezed field. Moreover, we show that it is a good way to enhance entanglement and squeezing by increasing the nonlinear gain coefficient of squeezed driving. Our results denote that magnon-cavity QED system is a powerful platform for studying macroscopic quantum phenomena, which illustrates a new method to  photon-photon entanglement and magnon squeezing.\\
\end{abstract}


\end{frontmatter}

%
%
%


	
\section{Introduction}

In recent years, yttrium iron garnet (YIG) material, as an excellent ferrimagnetic material with high spin density(about $4.22\times10^{-27}$ $\text m^{-3}$) and low dissipation rate(about 1 $\text {MHz}$), has attracted considerable attention~\cite{2019, flower2019broadening, zhu2020waveguide}. Moreover, YIG material is ferromagnetic at both cryogenic \cite{HueblZollitsch-609,TabuchiIshino-611,GoryachevFarr-545} and room temperature \cite{ZhangZou-2438} because its Curie temperature is about 559K. The magnon mode, as a collective motion of a large number of spins with zero wavevector (Kittel mode~\cite{Kittel-1121}) via the Holstein-Primakoff transformation~\cite{PrimakoffHolstein-1184} in YIG sphere, possesses unique properties. It can realize strong~\cite{HueblZollitsch-609,TabuchiIshino-611,GoryachevFarr-545,ZhangZou-2438,BaiHarder-614,ZhangWang-460} and ultrastrong~\cite{BourhillKostylev-561,KostylevGoryachev-2232} coupling to microwave cavity photons at either cryogenic or room temperature, and then lead to magnon-cavity polaritons. Thus, a lot of meaningful development about magnons is found, including the observation of cavity spintronics~\cite{BaiHarder-614,BaiHarder-619}, bistability~\cite{WangZhang-1124}, magnon gradient memory\cite{ZhangZou-324}, magnetically controllable slow light~\cite{KongWang-463}, level attraction\cite{HarderYang-624}, magnon-induced transparency~\cite{WangLiu-1171,ZhangZou-648}, and the magnon squeezed state~\cite{LiZhu-538}. It is noted that magnon squeezed state is an important macroscopic quantum state, which can be used to improve the measurement sensitivity~\cite{ebrahimi2021single} and study decoherence theories at large scales\cite{BassiLochan-2448}.

Meanwhile, by virtue of strong coupling among magnons, other interesting phenomena, including coupling the magnon mode to a single superconducting qubit~\cite{TabuchiIshino-2226}, to photons~ and to phonon mode~\cite{ZhangZou-648,ZhangWang-2451}, have also been studied. This offers a possibility to enable coherent information transfer between different information carriers. Clearly, compared to atom, the size of YIG sphere is in mesoscopic or macroscopic scale usually with a size of $\sim250\mu m$, which holds the potentiality for implementing quantum states, especially the entanglement in more massive object. Thus, it provides a promising and completely new platform for the study of macroscopic quantum phenomena~\cite{ZhangWang-2451}, which is a key step to test decoherence theories at macroscopic scale~\cite{BassiLochan-2448,WeaverNewsom-2450}, and probe the boundary between the quantum and classical worlds~\cite{Chen-2456,Leggett-2455,FroewisSekatski-2457}. In the microwave region, one important quantum state is entangled state, which is typically produced by exploiting the nonlinearity of magnetostrictive interaction in cavity magnomechanical system\cite{YuShen-2341}, by utilizing Kerr nonlinearity results from magnetocrystalline anisotropy\cite{ZhangScully-631}, and by using the nonlinearity of quantum noise in Josephson parametric amplifiers (JPA) \cite{NairAgarwal-2024}. Meanwhile, another important macroscopic quantum state is magnon squeezed state, which is usually generated by the quantum noise of JPA process\cite{LiZhu-538}.

Recent interest has focused on generating entanglement and squeezing in a hybrid cavity magnon QED system, especially in a hybrid cavity magnomechanics system including phonons. A genuine tripartite entanglement is shown by using the nonlinearity of magnon-phonon coupling in a cavity magnomechanical system consisting of magnons, microwave photons and phonons~\cite{LiZhu-1108}, where the magnons couple to microwave photons and phonons via magnetic dipole interaction and magnetostrictive interaction, respectively. When driving the above cavity (Ref.~\cite{LiZhu-1108}) by a weak squeezed vacuum field generated by a flux-driven JPA process, the magnons and phonons are squeezed in succession, and larger squeezing could be realized by increasing the degree of squeezing of the drive field and working at a lower temperature \cite{LiZhu-538}. A hybrid cavity magnomechanical system includes two magnon modes in two macroscopic YIG spheres, which couple to the single microwave cavity mode via magnetic dipole interaction. By activating the nonlinear magnetostrictive interaction in one YIG sphere, realized by driving the magnon mode with a strong red-detuned microwave field, the two magnon modes get entangled~\cite{LiZhu-339}. When two YIG spheres are placed inside two microwave cavities driven by a two-mode squeezed microwave field. Each magnon mode couples to the cavity mode via magnetic dipole interaction. The quantum correlation of the two driving fields can efficiently transferred to the two magnon modes and magnon-magnon entanglement can be achieved. The two cavity modes also can entangle to each other~\cite{YuZhu-2359}. When considering the vibrational modes in the above cavity (Ref.~\cite{YuZhu-2359}), each phonon mode couples to the magnon mode via magnetostrictive interaction. By directly driving magnon mode with a strong red-detuned microwave field to active the magnomechanical anti-stokes process, and further driving the two cavities by a two-mode squeezed vacuum field as above scheme (Ref.~\cite{YuZhu-2359}), the two phonon modes in two YIG spheres can also get entangled~\cite{LiGroblacher-2444}. All the above solutions denote that magnetic dipole coupling interaction and nonlinearity are two main elements to produce entangled and squeezed states. The main nonlinearity used is the nonlinearity of magnetostrictive interaction\cite{LiZhu-339,YuShen-2341,ZhangWang-2451,LiZhu-1108} and the nonlinearity generated by quantum noise of JPA process\cite{YuZhu-2359,NairAgarwal-2024} in cavity magnon system.

In this letter, we propose a scheme to generate photon-photon entanglement and squeeze magnon in a magnon-cavity QED system. Magnon mode in a YIG sphere is coupled to two microwave fields via magnetic dipole interaction, respectively. Since YIG material that generates magnons is a massive object, it is considered to be a theoretically innovation to realize the entanglement of two mesoscopic objects through a cavity mode. However, for the cavity mode or photon being a good carrier of information, we hope to entangle the two cavity modes with the magnons as a mesoscopic medium, and we think this is more important from the perspective of information. Squeezing is also a very important quantum resource, and we then emphasize the squeezing of magnons. It is found that squeezing can be transmitted to various objects in this QED system. Different from previous propose, the nonlinearity we used is generated by parametric down-conversion of JPA process. The intensity of that is flexible tunable, resulting in a squeezed cavity mode. Meanwhile, for the phonon mode can provide another non-linearity and make the system into a more complex one, we did not take it into consideration. In our QED system, entanglement transferred from magnon-cavity $a_1$ subsystem to magnon-cavity $a_2$ subsystem, and then transferred to two cavity modes subsystem. A genuinely tripartite entangled state is formed. Meanwhile, the squeezed driving also prepares the magnon mode in a squeezing state. Further, we show that increasing the nonlinearity gain coefficient of squeezed driving is a good way to enhance entanglement and squeezed. Moreover, we show that the optimal entanglement and squeezing generated when the coupling rates between the two cavity modes and magnon mode are the same.

\section{The model}

We consider a hybrid magnon-cavity QED system, which consists of two microwave cavity modes and a magnon mode, as depicted in Fig.1. A squeezed microwave cavity $1$ (with frequency $\omega_1$) is implemented by parametric down-conversion in JPA process. We assume that the nonlinear gain coefficient of JPA is $\Omega$. The second microwave cavity (with frequency $\omega_2$) is perpendicular to the microwave cavity $1$ without any non-linearity driving, the resonance frequency $\omega_2$ is close to that of cavity mode $a_1$. To achieve strong couplings between the YIG sphere and these two cavity modes, we place the YIG sample at the center of both cavities. Meanwhile, the magnons are quasiparticles, a collective motion of a large number of spins spatially uniform mode (Kittel mode~\cite{Kittel-1121}) in a massive YIG sphere. The magnetic field of cavity mode $a_1$ and $a_2$ are along $x$ and $y$ direction, respectively. The bias magnetic field $H$ is along $z$-axis for producing the Kittel mode. Strongly coupled is implemented via magnetic dipole interaction. Moreover, a microwave field with angular frequency $\omega_0$ and Rabi frequency $\varepsilon_p$ is applied along the $x$ direction to driving $a_1$. We assume the size of the YIG sample to be much smaller than the microwave wavelengths in our QED system, so the radiation pressure on YIG sample induced by microwave fields can be neglected. The Hamiltonian of the system reads

{\setlength\arraycolsep{0pt}
\begin{eqnarray}
	H/\hbar&&=\sum_{j=1,2}\omega_j a_j^{\dag} a_j+\omega_mm^{\dag}m+\sum_{j=1,2}g_j(a_j^{\dag}+a_j)(m+m^{\dag})\nonumber\\
	&&+\varepsilon_p(a_1e^{i\omega_{0}t}+a_1^{\dag}e^{-i\omega_{0}t})
	+\Omega(a_1^{2}e^{2i\omega_{0}t}+{a_1^{\dag}}^{2}e^{-2i\omega_{0}t})
\end{eqnarray}

where $a_j$ and $a_j^{\dag}$ are, respectively, the annihilation and creation operators of cavity mode j. $m(m^{\dag})$ is annihilation (creation) operator of magnon mode~\cite{StancilPrabhakar-206}, which represent the collective motion of spins via the Holstein-Primakoff transformation~\cite{PrimakoffHolstein-1184} in terms of Bosons, satisfying $[O,O^{\dag}]=1$ $(O=a_1, a_2, m)$. $\omega_{j}$ $(j=1,2)$ and $\omega_{m}$ present the resonance frequency of cavity modes $a_j$ and magnon mode, respectively. The frequency of magnon mode can be adjusted by the external bias magnetic field $H$ via $\omega_{m}=\gamma H$, where $\gamma/2\pi=28$GHz/T is the gyromagnetic ratio. $g_j$ denotes the linear coupling rate between magnon mode and cavity mode $a_j$, which currently can be (much) larger than the dissipation rates $\kappa_j$ and $\kappa_m$ of cavity mode $a_j$ and magnon mode, i.e. $g_j$ $>\kappa_j$, $\kappa_m$ $(j=1,2)$. It denotes the magnon-cavity QED system is in the strong coupling regime, but not in the ultrastrong coupling regime, and the rotating-wave approximation can be applied for the magnon-cavity interaction terms in our magnon-cavity QED system.
\captionsetup[figure]{labelfont={bf},name={Fig.},labelsep=period}	
\begin{figure}[htbp]
	\centering
	\includegraphics[width=8cm]{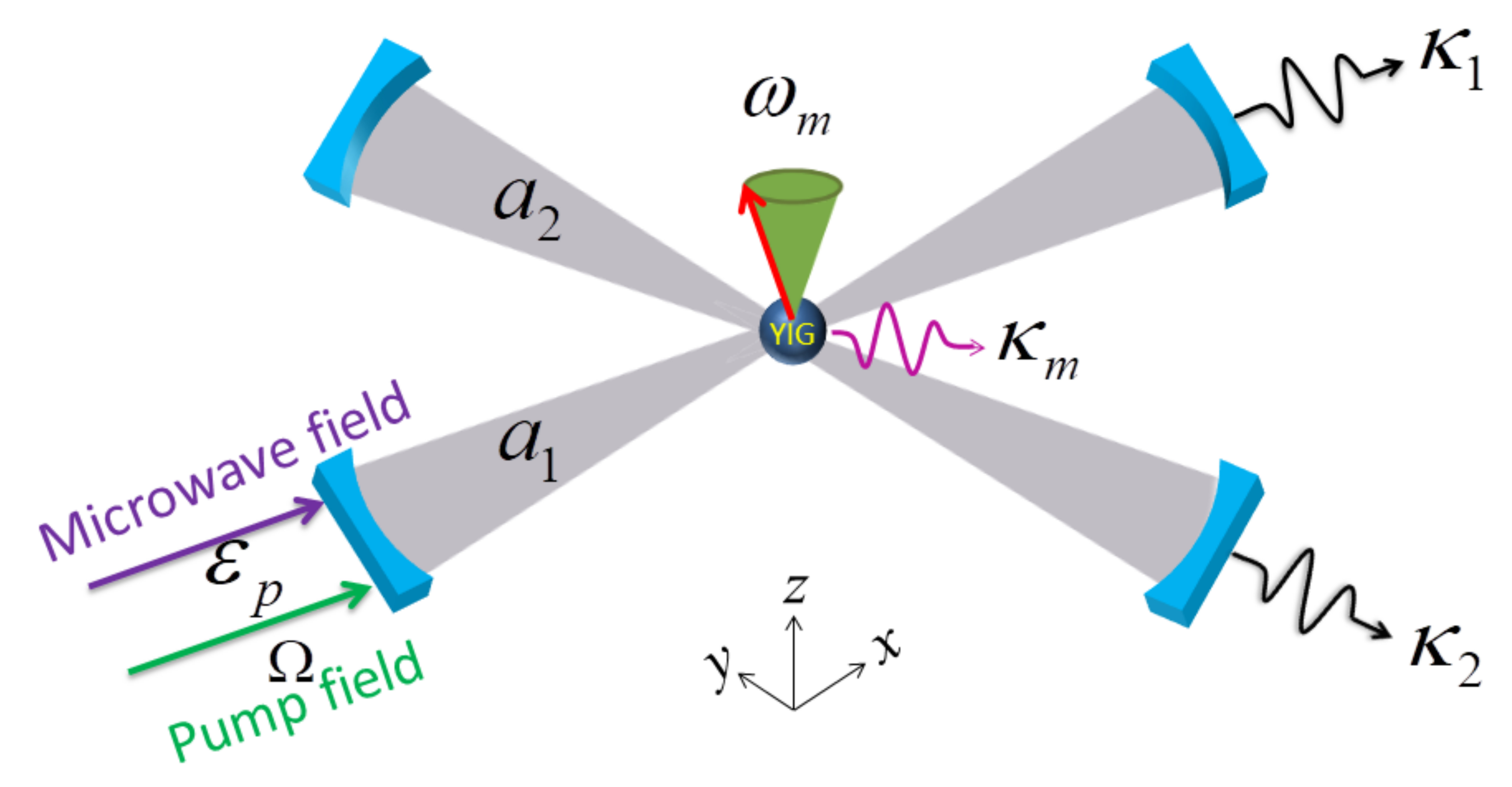}
	\caption{Schematic of magnon-cavity QED system. The first cavity is driven by a microwave field with $\varepsilon_p$ the Rabi frequency and a squeezed field with $\Omega$ the gain coefficient of parametric down-conversion, the resonance frequency of which is $\omega_1$. The second cavity (with frequency $\omega_2$) is perpendicular to the first one with a close angular frequency. The magnetic field of cavity mode $a_1$ and $a_2$ are along $x$ and $y$ direction, respectively. A YIG sphere is mounted at the center of the both microwave cavities. Simultaneously, it is also in a bias magnetic field $H$ along $z$-axis for producing the Kittel mode, resulting in the resonance frequency $\omega_{m}$. Here, $\kappa_{1}$, $\kappa_{2}$ and $\kappa_m$ are the dissipation rates of cavity mode $a_1$, cavity mode $a_2$ and magnon mode, respectively.}
	\label{Fig1exp}
\end{figure}

Under the rotating-wave approximation, the magnon-photon interaction term $g_{j}(a_j+a_j^{\dag})(m+m^{\dag})$ becomes $g_{j}(a_jm^{\dag}+a_j^{\dag}m)$. We then switch to the rotating frame with respect to the driving frequency $\omega_0$, the Hamiltonian of the system can be written as:

{\setlength\arraycolsep{0pt}
\begin{eqnarray}
	H/\hbar&=&\sum_{j=1,2}\Delta_j a_j^{\dag} a_j+\Delta_mm^{\dag}m+\sum_{j=1,2}g_j(a_j^{\dag}m+a_jm^{\dag})\nonumber\\
	&&+\varepsilon_p(a_1+a_1^{\dag})+\Omega(a_1^{2}+{a_1^{\dag}}^{2})
\end{eqnarray}

Where $\Delta_{j}=\omega_{j}-\omega_{0}$, and $\Delta_{m}=\omega_{m}-\omega_{0}$ are the detunings of cavity mode j and magnon mode, respectively. By including input noises and dissipations of the system, the quantum Langevin equations describing the system are as follows,
\begin{eqnarray}
	\dot{a_1}&=&-(i \Delta_1+\kappa_1) a_{1}-ig_{1}m-i\varepsilon_{p}-2i\Omega a_1^{\dag}+\sqrt{2\kappa_{1}}a_1^{in}\\
	\dot{a}_{2}&=&-(i\Delta_{2}+\kappa_{2})a_2-ig_{2}m+\sqrt{2\kappa_{2}}a_{2}^{in}\\
	\dot{m}&=&-(i\Delta_{m}+\kappa_{m})m-ig_1a_1-ig_{2}a_2+\sqrt{2\kappa_{m}}m^{in}
\end{eqnarray}

Where $a_j^{in}$ and $m^{in}$ are input noise operators for the cavity mode $a_j$ and magnon mode $m$, respectively, which are zero mean value acting on the cavity and magnon modes. The Gaussian nature of quantum noises can be characterized by the following correlation function~\cite{GardinerZoller-2453}:
$\langle{a_j^{in}(t)a_j^{in\dag}(t^{'})}\rangle=[N_j(\omega_{j})+1]\delta(t-t^{'})$,
$\langle{a_j^{in\dag}(t)a_j^{in}(t^{'})}\rangle=N_j(\omega_{j})\delta(t-t^{'})$$(j=1,2)$,
and $\langle{m^{in}(t)m^{in\dag}(t^{'})}\rangle=[N_{m}(\omega_{m})+1]\delta(t-t^{'})$,
$\langle{m^{in\dag}(t)m^{in}(t^{'})}\rangle=N_{m}(\omega_{m})\delta(t-t^{'})$
where $N_{l}(\omega_{l})=[exp(\hbar\omega_{l}/k_{B}T)-1]^{-1}(l=1,2,m)$ are the equilibrium mean thermal photon numbers and magnon number, respectively, with $k_B$ the Boltzmann constant and $T$ the environmental temperature.

Since the first cavity is under strong driving by the microwave field $\varepsilon_{p}$ and squeezed field $\Omega$, which results in a large amplitude $|\langle a_1\rangle|\gg1$ at the steady state. Meanwhile, due to the beam-splitter-like coupling interaction between cavity modes and magnon mode, magnon mode and cavity mode $a_2$ are also of large amplitudes in steady state. This allows us to linearize the system dynamics around the semiclassical averages and write any mode operator as $O=\langle O\rangle+\delta O(O=a_1,a_2,m)$, neglecting small second-order fluctuation terms. Here, $\langle O\rangle$ is the mean value of the operator $O$, and $\delta O$ is the zero-mean quantum fluctuation. We then obtain two sets of equations for semiclassical averages and for quantum fluctuations. The former set of equations are given by:
\begin{eqnarray}
	&-(i& \Delta_1+\kappa_1) \langle a_{1}\rangle-ig_{1}\langle m\rangle-i\varepsilon_{p}-2i\Omega \langle a_1^{\dag}\rangle=0\\
	&-(i&\Delta_{2}+\kappa_{2})\langle a_2\rangle-ig_{2}\langle m\rangle=0\\
	&-(i&\Delta_{m}+\kappa_{m})\langle m\rangle-ig_1\langle a_1\rangle-ig_{2}\langle a_2\rangle=0
\end{eqnarray}
By solving Eqs.(6)-(8), we obtain the steady-state solution for the average values
\begin{eqnarray}
	&&\langle a_1\rangle=\frac{\frac{2 \Omega \varepsilon_{p}}{P}-\varepsilon_{p}}{\Delta_{1}-i\kappa_1-\frac{4 \Omega^{2}}{P}-\frac{g_{1}^{2}(\Delta_{2}-i \kappa_{2})}{(\Delta_{m}-i \kappa_{m})(\Delta_2-i\kappa_2)-g_2^2}}\\
	&&\langle a_{2}\rangle=\frac{g_1g_{2}\langle a_1\rangle}{(\Delta_{m}-i \kappa_{m})(\Delta_2-i\kappa_2)-g_2^2}\\
	&&\langle m\rangle=\frac{-g_1(\Delta_2-i\kappa_2)\langle a_1\rangle}{(\Delta_{m}-i \kappa_{m})(\Delta_2-i\kappa_2)-g_2^2}
\end{eqnarray}
where $P=\Delta_{1}+i\kappa_1-\frac{g_{1}^{2}(\Delta_2+i\kappa_2)}
{(\Delta_{m}+i \kappa_{m})(\Delta_2+i\kappa_2)-g_{2}^{2}}$. Thus, we can obtain the mean photon numbers and mean magnon number from Eqs.(9)-(11).

On the other hand, quantum fluctuations is related to entanglement and squeezing. To study the quantum characteristics of the two cavity modes and magnon mode, the quadratures of quantum fluctuations about cavity modes and magnon mode are as
$\delta X_1=(\delta a_1+\delta a_1^{\dag})/\sqrt2$,
$\delta Y_1=i(\delta a_1^{\dag}-\delta a_1)/\sqrt2$,
$\delta X_2=(\delta a_2+\delta a_2^{\dag})/\sqrt2$,
$\delta Y_2=i(\delta a_2^{\dag}-\delta a_2)/\sqrt2$,
$\delta x=(\delta m+\delta m^{\dag})/\sqrt2$,
and $\delta y=i(\delta m^{\dag}-\delta m)/\sqrt2$,
and similarly for the input noise operators.
The quantum Langevin equations describing quadrature fluctuations
$(\delta X_1,$
$\delta Y_1,$
$\delta X_2,$
$\delta Y_2,$
$\delta x,$
$\delta y)$
can be written as
\begin{equation}
	\dot{f}(t)=Af(t)+\eta
\end{equation}
where $f(t)=[\delta X_1(t)$,$\delta Y_1(t)$,$\delta X_2(t)$,$\delta Y_2(t)$,$\delta x(t)$,$\delta y(t)]^{T}$ and
$\eta(t)=[\sqrt{2\kappa_1}X_1^{in}(t)$,
$\sqrt{2\kappa_1}Y_1^{in}(t)$,
$\sqrt{2\kappa_2}X_2^{in}(t)$,
$\sqrt{2\kappa_2}Y_2^{in}(t)$,
$\sqrt{2\kappa_m}x^{in}(t)$,
$\sqrt{2\kappa_m}y^{in}(t)]^{T}$ are the vectors for quantum fluctuations operator and noises operator, respectively. The drift matrix A is given by

$A=\left(\begin{array}{cccccc}{-\kappa_1}&{\Delta_{1}-2\Omega}&{0}&{0}&{0}&{g_{1}}\\
	{-\Delta_{1}-2\Omega}&{-\kappa_1}&{0}&{0}&{-g_{1}}&{0}\\
	{0}&{0}&{-\kappa_{2}}&{\Delta_{2}}&{0}&{g_{2}}\\
	{0}&{0}&{-\Delta_{2}}&{-\kappa_{2}}&{-g_{2}}&{0}\\
	{0}&{g_{1}}&{0}&{g_{2}}&{-\kappa_m}&{\Delta_{m}}\\ {-g_{1}}&{0}&{-g_{2}}&{0}&{-\Delta_{m}}&{-\kappa_{m}}\end{array}\right)$\\

Due to the linearized dynamics and the Gaussian nature of the quantum noises in our system, the steady state of quantum fluctuations is a continuous variable three mode Gaussian state, which is completely characterized by a $6\times6$ covariance matrix $V$ defined as
$V_{ij}=\langle f_i(t)f_j(t^{'})+f_j(t^{'})f_i(t)\rangle/2$ $(i,j=1,2,...,6)$.
In generally, the steady-state covariance matrix $V$ can be obtained straightforwardly by solving the Lyapunov equation~\cite{VitaliGigan-2452,ParksHahn-299}
\begin{equation}
	AV+VA^{T}=-D
\end{equation}
where $D=$diag$[\kappa_{1}\left(2N_{1}+1\right)$,
$\kappa_{1}\left(2N_{1}+1\right)$,
$\kappa_{2}\left(2N_{2}+1\right)$,
$\kappa_{2}\left(2N_{2}+1\right)$,
$\kappa_{m}\left(2N_{m}+1\right)$,
$\kappa_{m}\left(2N_{m}+1\right)]$ is the diffusion matrix, which is defined as $D_{ij}\delta(t-t^{'})=\langle \eta_i(t)\eta_j(t^{'})+\eta_j(t^{'})\eta_i(t)\rangle/2$. With the covariance matrix in hand, we can get the quantities related to entanglement and squeezing. To quantify entanglement between the two cavity modes and magnon mode, we adopt quantitative measures of the logarithmic negativity~\cite{VidalWerner-490,Plenio-591} $E_N$ for the bipartite entanglement, which is defined as
\begin{equation}
	E_N\equiv max[0,-ln2\tilde{\nu}_-]
\end{equation}
where $\tilde{\nu}_{-}=$min$[$eig$|i\Omega_2\tilde{V}_{4}|]$ is the minimum symplectic eigenvalue of the $\tilde{V}_{4}=P_{1\mid2}V_4P_{1\mid2}$. $V_4$ is the $4\times 4$ covariance matrix, which can be obtained by directly removing in $V$ the rows and columns of uninteresting mode. Meanwhile, to realize partial transposition at the level of covariance matrix, we set $P_{1\mid2}=$diag$(1,-1,1,1)$. $\Omega_2$ is symplectic matrix with $\Omega_2=\oplus^2_{j=1}i\sigma_y$ and $\sigma_y$ is the $y$-Pauli matrix. A nonzero logarithmic negativity $E_N>0$ denotes the presence of bipartite entanglement in our QED system.

Meanwhile, a quantification of continuous variable tripartite entanglement is given by the minimum residual contangle~\cite{AdessoIlluminati-1195,AdessoIlluminati-1196}, defined as
\begin{equation}
	R_{\tau}^{min}\equiv min[R_{\tau}^{a\mid m_1m_2},R_{\tau}^{m_1\mid am_2},R_{\tau}^{m_2\mid am_1}]
\end{equation}
where $R_{\tau}^{i\mid jk}\equiv C_{i\mid jk}-C_{i\mid j}-C_{i\mid k}\geq 0$ $(i,j,k=a,m_1,m_2)$ is the residual contangle, with $C_{u\mid v}$ the contangle of subsystems of $u$ and $v$ ($v$ contains one or two modes), which is a proper entanglement monotone defined as the squared logarithmic negativity. When $v$ contains two modes, logarithmic negativity $E_{i|jk}$ can be calculated by the definition of Eq.(14). We only need to use $\Omega_3=\oplus^3_{j=1}i\sigma_y$ instead of $\Omega_2=\oplus^2_{j=1}i\sigma_y$ and $\tilde{V}_{6}=P_{i\mid jk}VP_{i\mid jk}$ instead of $\tilde{V}_{4}=P_{1\mid2}V_4P_{1\mid2}$, where
$P_{1\mid 23}=diag(1,-1,1,1,1,1)$,
$P_{2\mid 13}=diag(1,1,1,-1,1,1)$
and $P_{3\mid 12}=diag(1,1,1,1,1,-1)$ are partial transposition matrices. $R_{\tau}^{min}\geq 0$ denotes the presence of genuine tripartite entanglement in three modes Gaussian system. 

Meanwhile, squeezing can be calculated by the covariance matrix of quantum fluctuations. The variances of squeezed magnon quadratures are amplitude quadrature $\langle\delta x(t)^2\rangle$, phase quadrature $\langle\delta y(t)^2\rangle$, and amplitude quadrature $\langle\delta Y_2(t)^2\rangle$ is quadrature of cavity mode $a_2$, $\delta x=(\delta m^\dag+\delta m)/\sqrt 2$, $\delta y=i(\delta m^\dag-\delta m)/\sqrt 2$, and $\delta Y_2=i(\delta a_2^\dag-\delta a_2)/\sqrt 2$. In our definition, $\langle\delta Q(t)^2\rangle _{\text vac}=1/2$ (Q is a mode quadrature) denotes vacuum fluctuations. The degree of squeezing can be expressed in the dB unit, which can be evaluated by $-10 log_{10}[\langle\delta Q(t)^2\rangle/\langle\delta Q(t)^2\rangle_{\text vac}]$, where $\langle\delta Q(t)^2\rangle _{\text vac} =1/2$.

\section{Results and discussion}

To show whether the squeezed driving can induce entanglement, we consider a simpler magnon-cavity QED system at first, where no coupling interaction exists between the magnon mode and cavity mode $a_2$, i.e., $g_2=0$. Fig.2(a) shows the bipartite entanglement between cavity mode $a_1$ and magnon mode versus detunings $\Delta_1$ and $\Delta_m$ in steady state. We employed experimentally feasible parameter~\cite{TabuchiIshino-611} at low temperature $T=10$mK, as
$\omega_1/2\pi=10$GHz,
$\kappa_m/2\pi=1$MHz,
$\kappa_{1}/2\pi=\kappa_{2}/2\pi=5$MHz,
$g_{1}/2\pi=20$MHz.
Moreover, Rabi frequency of microwave field we employed is $\varepsilon_p=10\kappa_m$. Squeezed field used in our system is to generate nonlinear term by the JPA process with gain coefficient $\Omega=2.5\kappa_m$. This is the nonlinearity that causes entanglement in our QED system. Fig.2(a) shows that the photon-magnon entanglement described by logarithmic negativity can achieve to 0.3. Meanwhile, due to the state-swap interaction between the cavity mode $a_1$ and magnon mode, the squeezing can be transferred from squeezed cavity mode $a_1$ to the magnon mode, as shown in Fig.2(b).

Note that the above results are valid only when the assumption of low-lying excitations, i.e. magnon excitation number $\langle m^{\dag}m\rangle \ll 2Ns$, where $s=5/2$ is the spin number of ground-state $\text F\text e^{3+}$ ion in YIG sphere. The total number of spins $N=\rho V$ with $\rho=4.22\times10^{27}\text m^{-3}$ the spin density of YIG and $V$ the volume of sphere. For a 250-$\mu m$-diameter YIG sphere, the number of spins $N\simeq 3.5\times10^{16}$. We then calculate the mean photon numbers of cavity mode $a_1$ $N_1=\langle a_1^{\dag}a_1\rangle$, cavity mode $a_2$ $N_2=\langle a_2^{\dag}a_2\rangle$, and mean magnon number $N_m=\langle m^{\dag}m\rangle$ via Eqs.(9)-(11), which are closely related to the input intensity of microwave field and squeezed field. Fig.2(c) and (d) show the mean photon number $N_1$ and mean magnon number $N_m$ versus detunings $\Delta_1$ and $\Delta_m$ in steady state when $g_2=0$. They are drawn with logarithmic $log_{10}$. We show that both the maximum number of photons and magnons are above 10, but less than $10^3$ in Fig.2(c) and (d). Meanwhile, we also get $N_2=0$. so the assumption of low-lying excitations is well satisfied.

\begin{figure}[htbp]
	\centering
	\includegraphics[width=8cm]{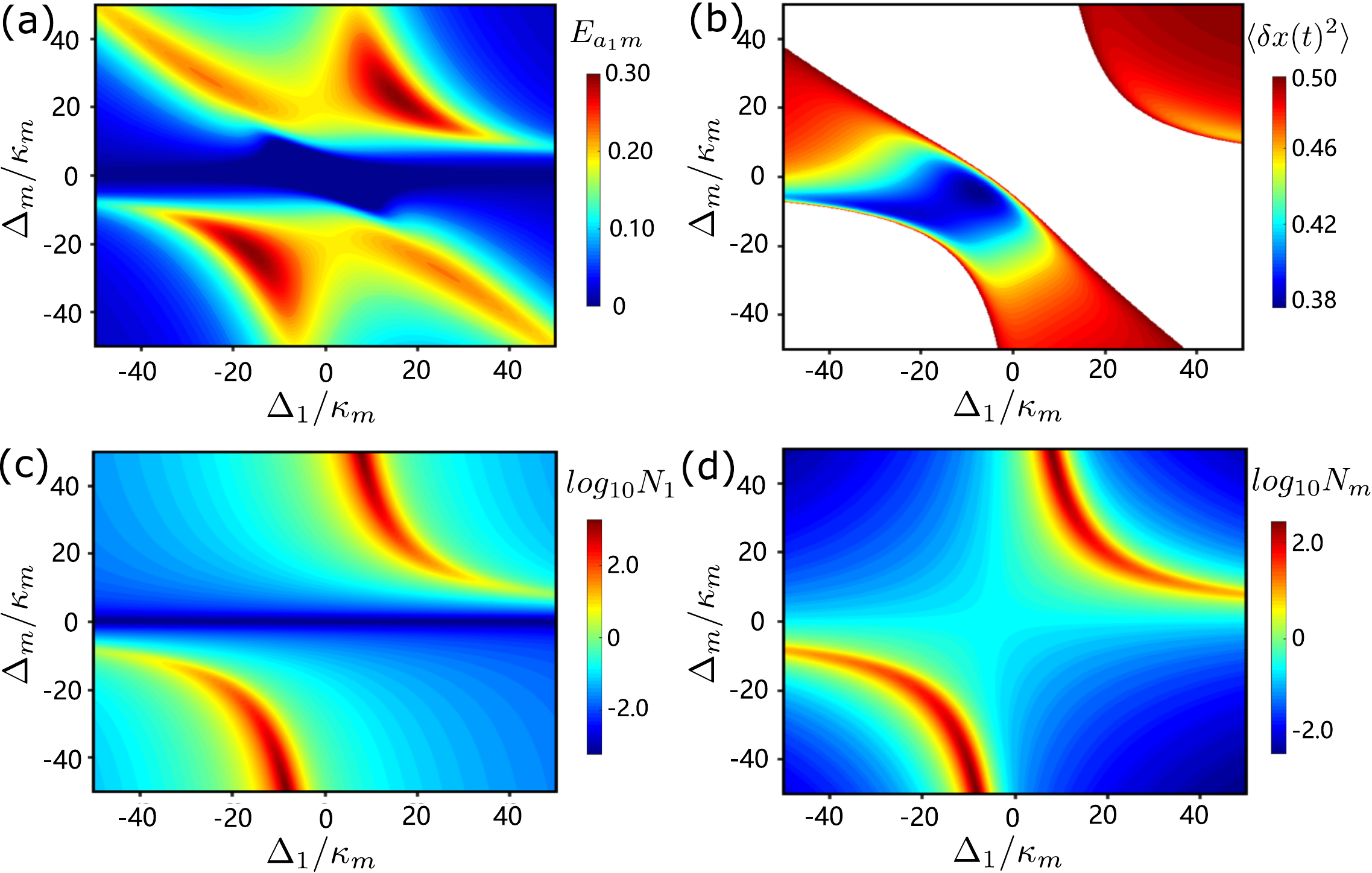}
	\caption{(a)Density plot of photon-magnon bipartite entanglement $E_{a_1m}$, (b)variance of the magnon amplitude quadrature $\langle\delta x(t)^2\rangle$, (c)logarithm of mean photon number of cavity mode $a_1$ $N_1$, and (d)logarithm of mean magnon number $N_m$ versus detunings $\Delta_1$ and $\Delta_{m}$. We choose $\Omega=2.5\kappa_m$, $\varepsilon_p=10\kappa_m$. The blank area denotes $\langle\delta Q(t)^2\rangle _{\text vac} >1/2$, i.e., above vacuum fluctuations. We take $g_2=0$ for all the plots. See text for the detail of other parameters.}
	\label{EN}
\end{figure}

We then take $g_2$ into consideration. To be more general, we assume that coupling rate $g_2$ is the same as that between the cavity mode $a_1$ and magnon mode, i.e., $g_2=g_1$. In Fig.3(a)-3(c), mean photon numbers and mean magnon number, $N_1$, $N_2$, and $N_m$, are plotted as functions of detunings $\Delta_2$ and $\Delta_{m}$, respectively. They are also drawn with logarithmic $log_{10}$. It is noted that $P=0$ is the extreme value of Eqs.(9)-(11). Ignoring dissipative terms and analyzing the extreme value, we can obtain a simple form $\Delta_m=(\Delta_1g_2^2+\Delta_2g_1^2)/(\Delta_1\Delta_2)$. The black dashed curves in Fig.3(a)-(c) denote $\Delta_m=(\Delta_1g_2^2+\Delta_2g_1^2)/(\Delta_1\Delta_2)$, and from which we can see that the maximum numbers of photons and magnons are located at about this region.

\begin{figure}[htbp]
	\centering
	\includegraphics[width=8cm]{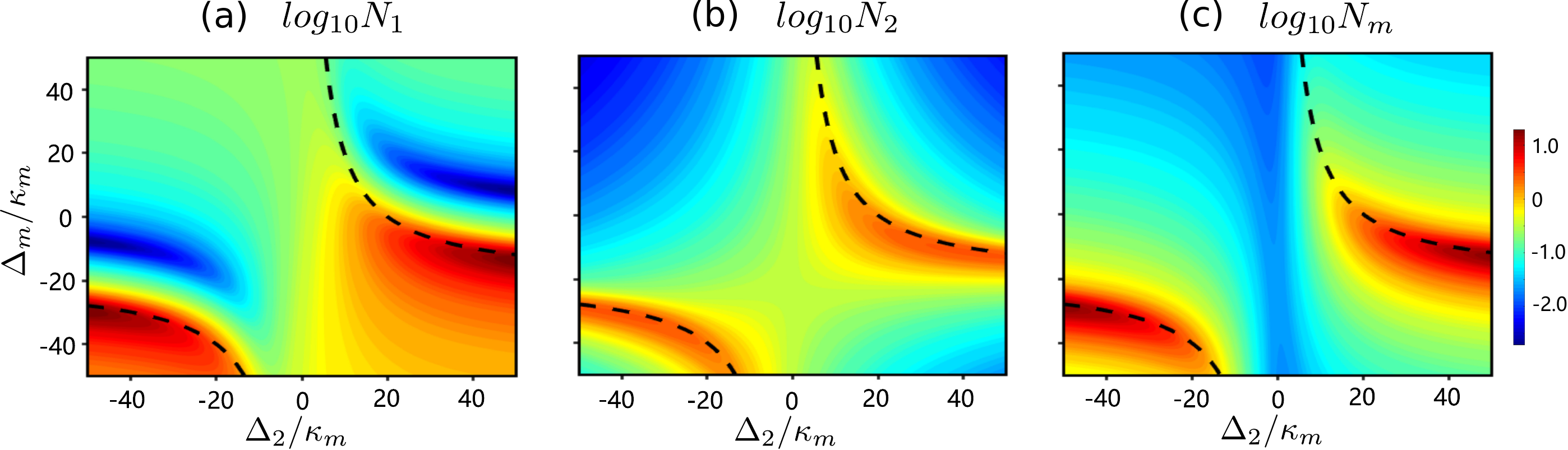}
	\caption{(a)Mean photon number of cavity mode $a_1$ (squeezed cavity mode) $N_1$, (b)mean photon number of cavity mode $a_2$ $N_2$, and (c)mean magnon number $N_m$ versus detunings $\Delta_2$ and $\Delta_{m}$. Black dash curves indicate $\Delta_m=(\Delta_1g_2^2+\Delta_2g_1^2)/(\Delta_1\Delta_2)$. All Figures are drawn with logarithmic $log_{10}$. We take Rabi frequency of microwave field $\varepsilon_p=10\kappa_m$ and the nonlinear gain coefficient of squeezed field $\Omega=2.5\kappa_m$. We assume the coupling rate between the two cavity modes and magnon mode are the same, i.e., $g_2=g_1$. The detuning of cavity mode $a_1$ $\Delta_1=-20\kappa_m$. The other parameters are as in Fig.2.}
	\label{TE102OmegaT}
\end{figure}

After coupling cavity mode $a_2$ to magnon mode ($g_2>0)$, the magnon-cavity $a_1$ entanglement $E_{a_1m}$ decreased while cavity mode $a_2$ and magnon mode get entangled, as shown in Fig.4(a) and (c) with assuming $g_2=g_1$. It denotes that the quantum correlations can be transferred from magnon mode and cavity mode $a_1$ to magnon mode and cavity mode $a_2$. All results are in the steady state guaranteed by the negative eigenvalues (real parts) of the drift matrix A. We also choose the Rabi frequency of microwave field $\varepsilon_p=10\kappa_m$ and the gain coefficient $\Omega=2.5\kappa_m$. The Black dashed curves in Fig.4(a) and (c) denote $\Delta_m=(\Delta_1g_2^2+\Delta_2g_1^2)/(\Delta_1\Delta_2)$. It clearly shows that the optimal photon-magnon entanglement is achieved near the maximum of mean particle numbers.

We then calculated the squeezing by the covariance matrix of quantum fluctuations applying mean field approximation, and found that the cavity modes and the magnon mode can be squeezed. Compared with photons, it is more meaningful to study squeezed magnons, a mesoscopic object. Two quadratures of magnon mode are amplitude quadrature $\langle\delta x(t)^2\rangle$ and phase quadrature $\langle\delta y(t)^2\rangle$, these two quadratures also obey the uncertainty relationship, i.e., when the phase (amplitude) quadrature is squeezed, the amplitude (phase) quadrature will not be squeezed. That is, the squeezed of one quadrature is at the expense of increasing the other one. Variance of magnon amplitude quadrature $\langle\delta x(t)^2\rangle$ and phase quadrature $\langle\delta y(t)^2\rangle$ versus detunings $\Delta_2$ and $\Delta_m$ are shown in Fig.4(b) and (d), respectively. The blank area denotes above vacuum fluctuations, i.e., $\langle\delta Q(t)^2\rangle _{\text vac} >1/2$, $(Q=x,y)$.

\begin{figure}[htbp]
	\centering
	\includegraphics[width=8cm]{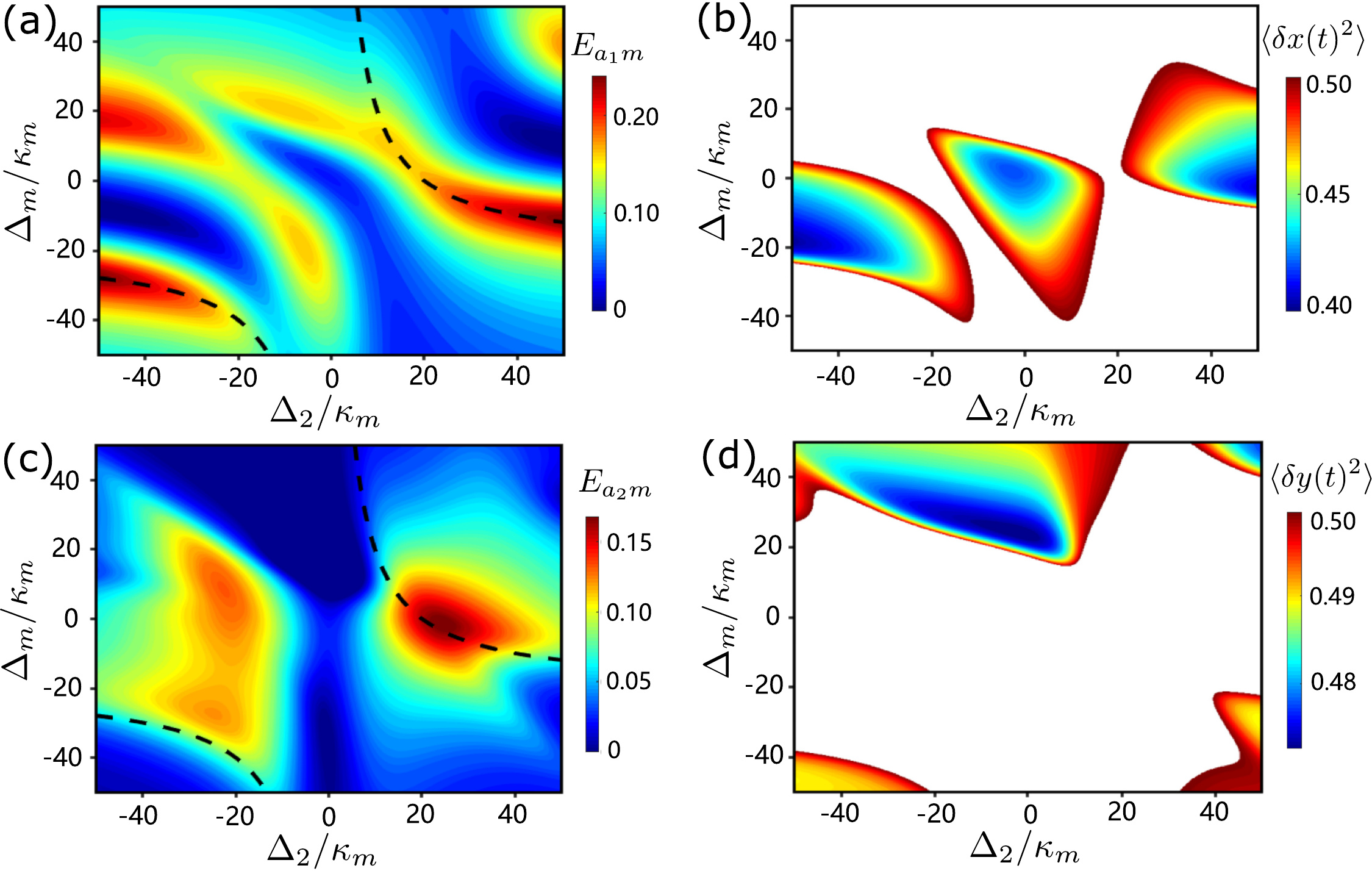}
	\caption{(a)Density plot of bipartite entanglement $E_{a_1m}$, (b)variance of the magnon amplitude quadrature $\langle\delta x(t)^2\rangle$, (c)density plot of bipartite entanglement $E_{a_2m}$, and (d)variance of the magnon phase quadrature $\langle\delta y(t)^2\rangle$ versus detunings $\Delta_2$ and $\Delta_{m}$. We choose $\Omega=2.5\kappa_m$, $\varepsilon_p=10\kappa_m$. The detuning of cavity mode $a_1$ $\Delta_1=-20\kappa_m$. Black dash curves in Fig.4(a) and (c) indicate $\Delta_m=(\Delta_1g_2^2+\Delta_2g_1^2)/(\Delta_1\Delta_2)$. The blank area in Fig.4(b) and (d) denotes $\langle\delta Q(t)^2\rangle _{\text vac} >1/2$, i.e., above vacuum fluctuation. We take $g_2=g_1$ for all the plots. See text for the other parameters.}
	\label{TE102R}
\end{figure}

Further, Fig.5(a) shows that the two cavity modes get entangled, which denotes that the photon-photon entanglement $E_{a_1a_2}$ is transferred from magnon-cavity entanglement $E_{a_1m}$ and $E_{a_2m}$  due to the state-swap interaction between the two cavity modes and magnon mode. The coupling rate $g_2$ also induces the squeezing transferred from cavity mode $a_1$ to cavity mode $a_2$ via magnon mode, as shown in Fig.5(b). The blank area denotes above vacuum fluctuations, i.e., $\langle\delta Q(t)^2\rangle _{\text vac} >1/2$. Comparing to Fig.2(b), the maximum of variance of the magnon amplitude quadratures $\langle \delta x(t)^2 \rangle $ and $\langle \delta y(t)^2 \rangle $ decreases, and cavity mode $a_2$ get squeezed. It denotes that the two cavity modes and magnon mode are all prepared in squeezed states due to the state-swap interaction between the two cavity modes and magnon mode, meaning that the magnetic dipole interaction is an essential element to generate squeezed states. Logarithmic negativity $E_{a_1a_2}$ as a function of bath temperature is shown in Fig.5(c). It denotes that photon-photon entanglement $E_{a_1a_2}$ is robust again bath temperature and survives up to about $200mK$. Tripartite entanglement in terms of the minimum residual contangle $R_{\tau}^{min}$ detunings $\Delta_2$ and $\Delta_{m}$ is shown in Fig5(d). It shows that the tripartite entanglement does exist in our QED system. The black dashed curves in Fig.5(d) denote $\Delta_m=(\Delta_1g_2^2+\Delta_2g_1^2)/(\Delta_1\Delta_2)$, and from which we can see that the maximum of tripartite entanglement located at about this region.

\begin{figure}[htbp]
	\centering
	\includegraphics[width=8cm]{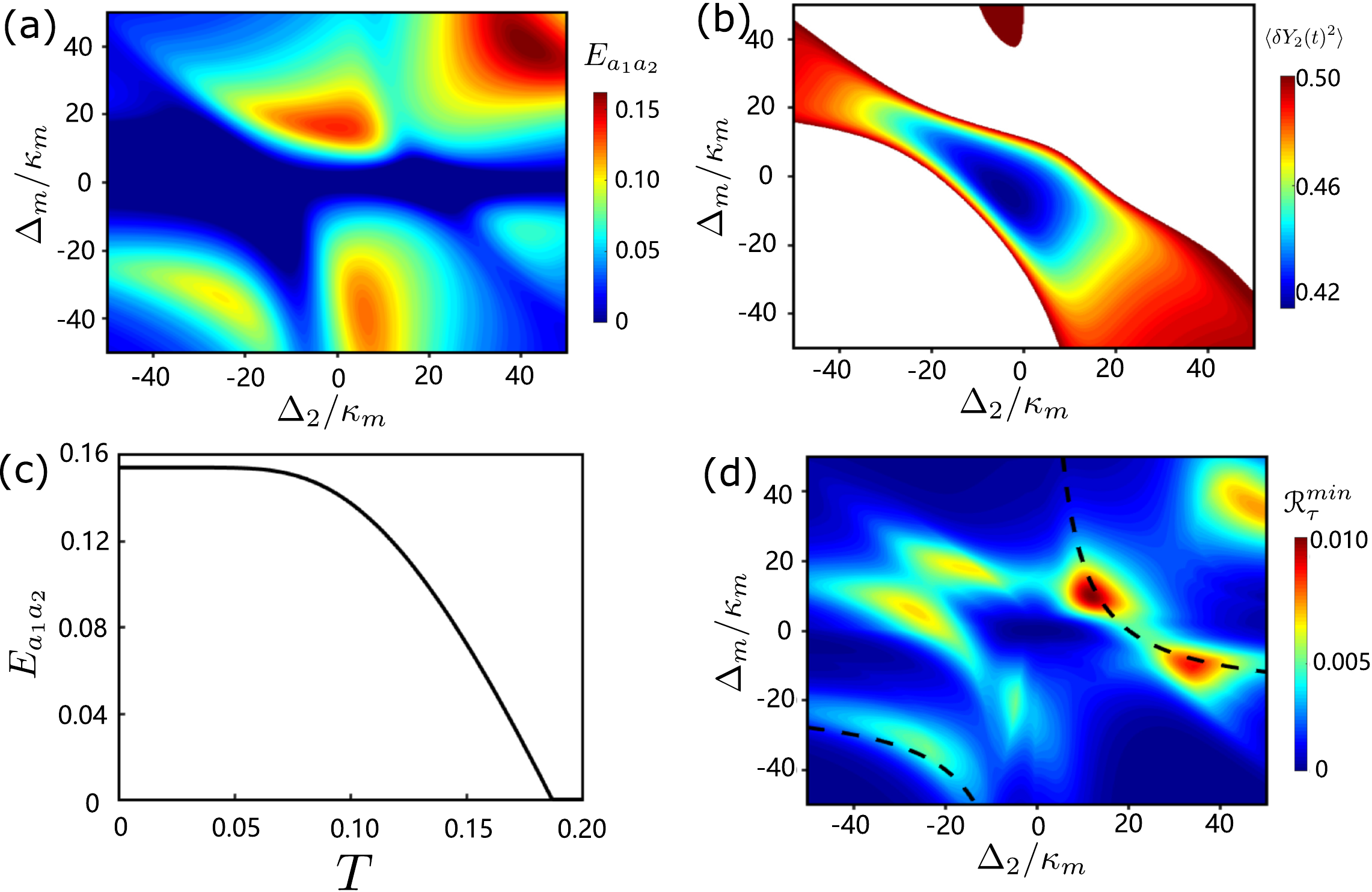}
	\caption{(a)Density plot of photon-photon bipartite entanglement $E_{a_1a_2}$, and (b)variance of cavity mode $a_2$ amplitude quadrature $\langle\delta Y_2(t)^2\rangle$ versus detunings $\Delta_2$ and $\Delta_{m}$. (c)Logarithmic negativity $E_{a_1a_2}$ vs bath temperature $T$. (d) Tripartite entanglement in terms of the minimum residual contangle $R_{\tau}^{min}$ detunings versus $\Delta_2$ and $\Delta_{m}$. The blank area in Fig.5(b) denotes $\langle\delta Q(t)^2\rangle _{\text vac} >1/2$, and black dash curves indicate $\Delta_m=(\Delta_1g_2^2+\Delta_2g_1^2)/(\Delta_1\Delta_2)$ in Fig.5(d). We take $\Delta_2=35\kappa_m$, $\Delta_m=45\kappa_m$ for (c), $\Omega=2.5\kappa_m$, $\varepsilon_p=10\kappa_m$, $\Delta_1=-20\kappa_m$ and $g_2=g_1$ for all the plots. See text for the other parameters.}
	\label{squzeed}
\end{figure}

Squeezing does not increase linearly with increasing the gain coefficient. We choose $\Delta_2=0$ , and find that squeezing first increases and then decreases with the increase of the gain coefficient, as shown in Fig.6(a) and (b), respectively.The blank area denotes $\langle\delta Q(t)^2\rangle _{\text vac} >1/2$. Squeezing reaches the maximum near $\Omega=8\kappa_m$ for the amplitude quadrature and near $\Omega=2\kappa_m$ for phase quadrature.

\begin{figure}[htbp]
	\centering
	\includegraphics[width=8cm]{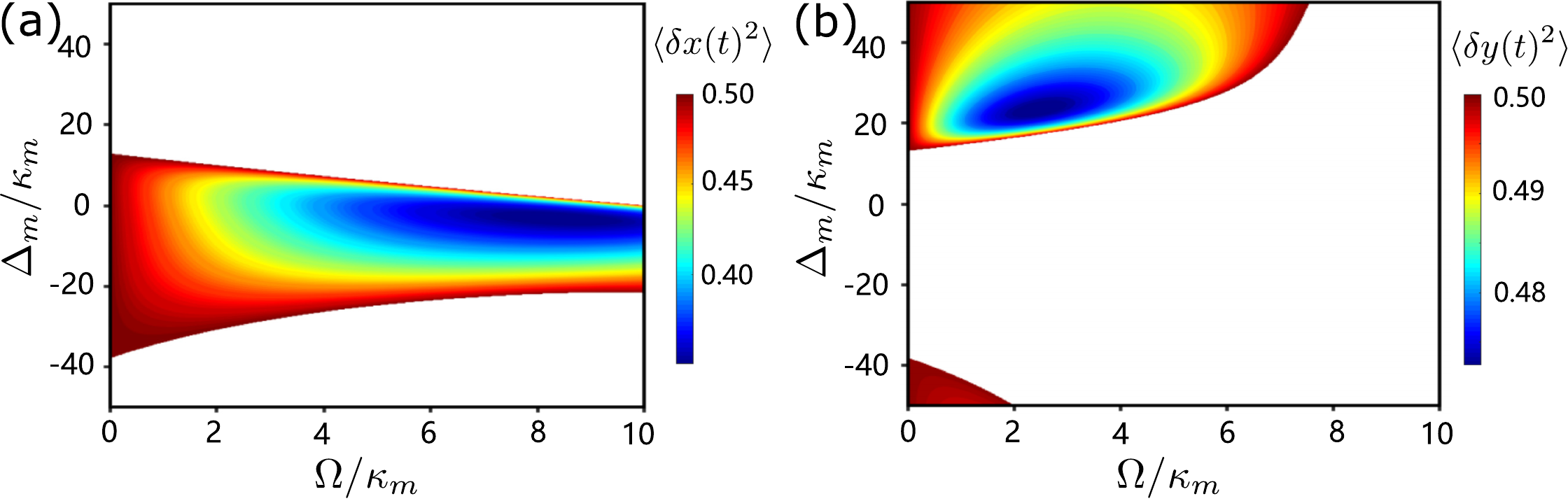}
	\caption{(a)Variance of the magnon amplitude quadrature $\langle \delta x(t)^2 \rangle$, (b)Variance of the magnon phase quadrature $\langle \delta y(t)^2 \rangle$ versus gain coefficient $\Omega$ and detunings $\Delta_{m}$. The blank area denotes $\langle\delta Q(t)^2\rangle _{\text vac} >1/2$. We take $\Delta_1=-20\kappa_m$, $\Delta_2=0$, $\varepsilon_p=10\kappa_m$ for all the plots. See text for the other parameters.}
	\label{squzeed}
\end{figure}

To obtain the optimal entanglement between the two cavity modes, we show photon-photon entanglement $E_{a_1a_2}$ versus gain coefficient $\Omega$ and the rate of magnon-cavity coupling strength $g_2/g_1$ in Fig.7(a). All results are calculated in the steady state, and the blank area denotes Non equilibrium state. As shown in Fig.7(a), the two cavity modes show zero entanglement in the absence of gain coefficient, i.e., $\Omega=0$, meaning that it is squeezed driving that induced entanglement in our QED system. It demonstrates that the nonlinearity produced by parametric down-conversion is the reason to generate entanglement. Bipartite entanglement $E_{a_1m}$ increases with the increase of gain coefficient $\Omega$, and then the entanglement transferred from $E_{a_1m}$ to $E_{a_2m}$ and $E_{a_1a_2}$. But, to keep the system in steady state, the gain coefficient can not be too large. Fig.7(a) denotes that increasing gain coefficient is a good way to improve entanglement in our QED system. Due to the flexible tunability of gain coefficient, which makes large entanglement possible. Further, we show that the optimal entanglement can be generated when the rate of photon-magnon coupling strength are almost the same, i.e., $g_2/g_1\approx 1$.

Meanwhile, we show variance of magnon amplitude quadrature $\langle \delta x(t)^2 \rangle$ versus gain coefficient $\Omega$ and the rate of magnon-cavity coupling strength $g_2/g_1$ in Fig.7(b). The blank area represents above vacuum fluctuation, i.e., \\
$\langle\delta Q(t)^2\rangle _{\text vac}>1/2$. It shows that the magnon mode can not be squeezed in the absence of squeezed field, i.e., $\Omega=0$, and the strength of squeezed magnon mode transferred from squeezed cavity mode $a_1$ can increase a lot as gain coefficient $\Omega$ increasing. It provides a good scheme to improve macroscopic quantum state. Further, we also show that the optimal squeezing is also located at about the region $g_2/g_1=1$.

\begin{figure}[htbp]
	\centering
	\includegraphics[width=8cm]{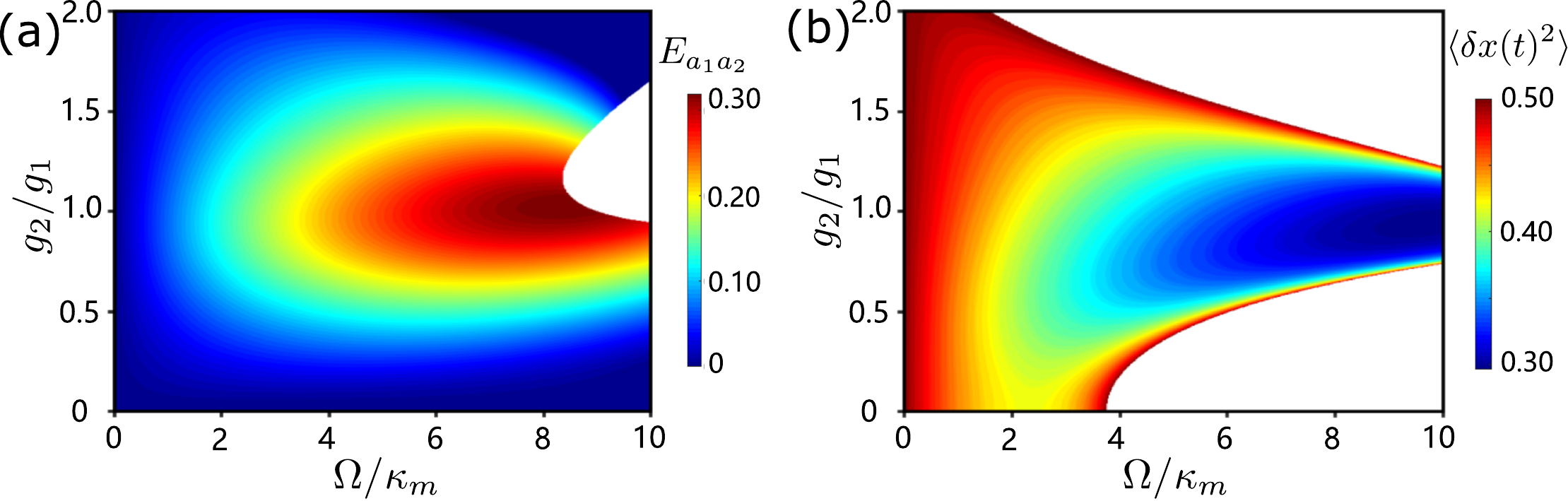}
	\caption{(a)Density plot of photon-photon bipartite entanglement $E_{a_1a_2}$, (b)variance of the magnon amplitude quadrature $\langle \delta x(t)^2 \rangle$ versus nonlinear gain coefficient $\Omega$ and the rate of magnon-cavity coupling strength $g_2/g_1$. The blank area in Fig.7(a) presents non equilibrium states and $\langle\delta Q(t)^2\rangle _{\text vac} >1/2$ in Fig.7(b), i.e., above vacuum fluctuations. We take $\Delta_2=35\kappa_m$, $\Delta_m=45\kappa_m$ for (a), $\Delta_2=-45\kappa_m$, $\Delta_m=-15\kappa_m$ for (b), and $\Delta_1=-20\kappa_m$, $\varepsilon_p=10\kappa_m$ for all the plots. See text for the other parameters.}
	\label{squzeed}
\end{figure}

\section{Conclusion}

In summary, we have presented a scheme to generate bipartite entanglement between two cavity modes and squeeze magnon mode in a magnon-cavity QED system by using a squeezed driving. With experimentally reachable parameters, we show that without the nonlinearity induced by parametric down-conversion process, our QED system denotes zero entanglement and above vacuum fluctuations. We also show the photon-magnon entanglement can transfer to photon-photon entanglement by state-swap interaction between cavity and magnon modes in the steady state. A genuinely tripartite entangled state is formed. Meanwhile, magnon squeezed state also can be realized due to the squeezing from squeezed driving cavity mode. Moreover, our QED system shows that increasing the nonlinear gain coefficient is a good way to enhance entanglement and squeezing. Further, the optimal entanglement and squeezing is located at about the region where the coupling rates between two cavity modes and magnon mode are almost the same. Our results denote that magnon-cavity QED system is a powerful platform for studying macroscopic quantum phenomena, and squeezed drive provides an new method for generating macroscopic quantum state.

\section*{Acknowledgements}
	This work has been supported by the National Natural Science Foundation of China (Grant No. 12174288, Grant No. 11874287, Grant No. 11774262, Grant No. 61975154), and the Shanghai Science and Technology Committee (Grant No. 18JC1410900).

%
%
%
%
%









%
%
%
\bibliographystyle{unsrt}

\bibliography{Squeezeddriving}
\end{document}